\documentclass[%
 reprint,
 preprintnumbers,
 amsmath,amssymb,
 aps,
]{revtex4-2}

\usepackage{graphicx}% Include figure files
\usepackage{dcolumn}% Align table columns on decimal point
\usepackage{bm}% bold math
\usepackage{hyperref}% add hypertext capabilities
\hypersetup{
	colorlinks=true,
	linkcolor=blue,
	filecolor=magenta,      
	urlcolor=blue,
	citecolor=magenta
}
\usepackage{amsmath,amssymb,mathtools}
\usepackage{graphics,graphicx,tikz,tensor}
\usetikzlibrary{arrows,decorations.pathmorphing,patterns}

\begin{document}

\preprint{UUITP-22/23}

\title{Angular Momentum Loss Due to Spin-Orbit Effects\\in the Post-Minkowskian Expansion}% Force line breaks with \\

\author{Carlo Heissenberg$^{\ast,\dagger}$}%\email{carlo.heissenberg@physics.uu.se}
\affiliation{$^\ast$Department of Physics and Astronomy, Uppsala University, Box 516, SE-75237 Uppsala, Sweden\\
	$^\dagger$NORDITA, KTH Royal Institute of Technology and Stockholm University,
	Hannes Alfv\'ens väg 12, SE-11419, Stockholm, Sweden}

%\collaboration{}%\noaffiliation


\begin{abstract}
We calculate the spin-orbit corrections to the loss of angular momentum in a two-body scattering at third Post-Minkowskian order, $\mathcal O(G^3)$, from scattering amplitudes using the eikonal operator. These results include effects linear in spin, are valid for generic spin orientations and are presented in a manifestly Poincar\'e covariant way. We include both  radiative losses, by means of the leading-order gravitational waveform, and static losses by means of the appropriate $-i0$ prescription in the leading soft graviton theorem, finding agreement with known results in the Post-Newtonian limit.
\end{abstract}

%Use showkeys class option if keyword                 

\maketitle

%\tableofcontents

\section{Introduction} 
Outstanding effort has been recently devoted to investigating gravitational observables associated to encounters of compact objects in the Post-Minkowskian (PM) approximation, which applies for weak gravitational fields and is valid for generic velocities, in particular leveraging the simplicity and on-shell nature of scattering amplitudes \cite{Bjerrum-Bohr:2018xdl,Kosower:2018adc,Bern:2019nnu,Bern:2019crd,Bern:2021dqo,Bern:2021yeh,Bern:2022jvn,Damgaard:2023vnx}.
The PM expansion lends itself more naturally to the study of unbound orbits, i.e.~scattering or hyperbolic-like encounters, in contrast with the Post-Newtonian (PN) expansion \cite{Blanchet:2013haa}, whereby one expands simultaneously for weak fields and small velocities. The latter is thus tied to bound or quasi-circular orbits, for which the virial theorem ensures that these two conditions must be satisfied simultaneously. Conversely, the relativistic nature of the PM expansion meshes well with the Lorentz invariance of scattering amplitudes, which makes exact results in the velocity easily accessible.

Besides improving the precision of the calculations, i.e.~including subleading PM orders, in order to produce phenomenologically accurate waveform templates \cite{Buonanno:2022pgc} it is of course important to capture various different physical effects, notably those due to spins \cite{Arkani-Hamed:2017jhn,Vines:2017hyw,Guevara:2018wpp,Chung:2018kqs,Maybee:2019jus,Guevara:2019fsj,Arkani-Hamed:2019ymq,Johansson:2019dnu,Chung:2019duq,Damgaard:2019lfh,Bautista:2019evw,Aoude:2020onz,Chung:2020rrz,Bern:2020buy,Guevara:2020xjx,Kosmopoulos:2021zoq,Aoude:2021oqj,Chiodaroli:2021eug,Haddad:2021znf,Chen:2021kxt,Aoude:2022trd,Bern:2022kto,Alessio:2022kwv,FebresCordero:2022jts,Cangemi:2022abk,Alessio:2023kgf,Bautista:2023szu,Aoude:2023vdk} and tides \cite{Cheung:2020sdj,Bern:2020uwk,Cheung:2020gbf,Aoude:2020ygw,AccettulliHuber:2020dal,Mougiakakos:2022sic,Heissenberg:2022tsn}. In particular, performing accurate parameter estimations of the colliding objects' spins and of their orientation can provide precious information concerning their formation mechanisms and astrophysical origin. For instance, binaries created by slow accretion of matter tend to have aligned spins, while  those arising form capture events tend to have misaligned spins \cite{Riva:2022fru}.

The bridge between scattering amplitudes and classical gravitational waves hinges on resummation methods like the eikonal exponentiation \cite{Kabat:1992tb,Akhoury:2013yua,KoemansCollado:2019ggb,Cheung:2020gyp,DiVecchia:2020ymx,AccettulliHuber:2020oou,DiVecchia:2021ndb,DiVecchia:2021bdo,Heissenberg:2021tzo,Bjerrum-Bohr:2021vuf,Bjerrum-Bohr:2021din,Damgaard:2021ipf,Brandhuber:2021eyq,DiVecchia:2022nna,DiVecchia:2023frv,Brandhuber:2023hhy,Georgoudis:2023lgf}, on effective-field-theory (EFT) techinques \cite{Goldberger:2004jt,Porto:2005ac,Porto:2016pyg,Cheung:2018wkq,Cristofoli:2020uzm} or on the observable-based KMOC framework \cite{Kosower:2018adc,Herrmann:2021lqe,Herrmann:2021tct,Cristofoli:2021vyo,Cristofoli:2021jas,Adamo:2022rmp,Adamo:2022qci,Herderschee:2023fxh,Elkhidir:2023dco,Damgaard:2023vnx,Damgaard:2023ttc}. 
Useful integration techniques borrowed from collider physics like differential equations \cite{Parra-Martinez:2020dzs,DiVecchia:2021bdo} and reverse unitarity \cite{Anastasiou:2002qz,Anastasiou:2002yz,Anastasiou:2003yy,Anastasiou:2015yha,Herrmann:2021lqe,Herrmann:2021tct} have also proven instrumental in reducing complicated  calculations in general relativity to tractable loop integrals. In fact, such techniques have been crucial both in the already mentioned amplitude calculations and in quantum-field-theory-inspired worldline setups that very efficiently encode the PM expansion \cite{Goldberger:2016iau,Kalin:2020mvi,Kalin:2020fhe,Kalin:2020lmz,Mogull:2020sak,Jakobsen:2021smu,Mougiakakos:2021ckm,Liu:2021zxr,Dlapa:2021npj,Jakobsen:2021lvp,Jakobsen:2021zvh,Riva:2021vnj,Dlapa:2021vgp,Jakobsen:2022fcj,Mougiakakos:2022sic,Jakobsen:2022psy,Kalin:2022hph,Dlapa:2022lmu,Jakobsen:2022zsx,Dlapa:2023hsl,Jakobsen:2023ndj}.

The eikonal operator framework \cite{Ciafaloni:2018uwe,Addazi:2019mjh,Damgaard:2021ipf,DiVecchia:2022nna,DiVecchia:2022owy,Cristofoli:2021vyo,DiVecchia:2022piu} can be regarded as a resummation of the KMOC formalism \cite{Kosower:2018adc} whereby the final state of the collision is obtained by applying a suitable exponential operator, built from scattering amplitudes, to the initial one. One then takes expectation values of observables associated to the scattering in this final state at the classical stationary point \cite{Cristofoli:2021jas,DiVecchia:2022piu}. This framework makes it particularly transparent how classical gravitational interactions and radiation emerge from the exchange or emission of large numbers of gravitons. Exchanges, which cause the scattering objects to deflect, are encoded in the c-number eikonal phase, while emissions take the form of a coherent superposition given by the exponential of the graviton's ladder operators contracted with the appropriate wave-shapes.

In this paper we consider dissipative effects induced by the (classical) spin of the scattering objects, focusing on the terms linear in spin, often termed spin-orbit in analogy with atomic systems. Our approach is based on the eikonal operator and on reverse unitarity. Using these tools, we first confirm the results of Ref.~\cite{Riva:2022fru} for the $O(G^3)$ radiated energy-momentum due to spin-orbit effects and obtain a new prediction: the angular momentum that the two-body system loses due to spin-orbit effects at $\mathcal O(G^3)$.
As is typically the case for angular momentum losses \cite{Manohar:2022dea,DiVecchia:2022owy,DiVecchia:2022piu,Heissenberg:2022tsn}, both emissions of dynamically propagating gravitational waves and static-field effects contribute to the final result.

Accurate expressions for energy and angular momentum losses are important, in combination with those of binding energies, as they serve as key ingredients of waveform models \cite{Damour:2008yg,Antonelli:2019ytb,Khalil:2022ylj,Buonanno:2022pgc} that are relevant for gravitational-wave data analysis, for instance by fixing the so-called radiation reaction force \cite{Manohar:2022dea}. 
In addition, the knowledge of the $O(G^3)$ energy and angular momentum loss enables one to deduce the linear radiation reaction terms in the $O(G^4)$ impulse \cite{Bini:2021gat,Manohar:2022dea,Jakobsen:2022zsx,Bini:2022enm,Jakobsen:2023hig}.

The paper is organized as follows. In Section~\ref{sec:general} we provide a concise presentation of the eikonal operator \cite{Cristofoli:2021jas,DiVecchia:2022piu,Heissenberg:2022tsn,DiVecchia:2023frv} and discuss the inclusion of spin in one of its key ingredients, the tree-level waveform, following in particular Ref.~\cite{Riva:2022fru}. In Section~\ref{sec:radiative} we then turn to the calculation of integrated radiative observables characterizing gravitational-wave emission: first energy-momentum, which confirms the result of Ref.~\cite{Riva:2022fru}, and then angular momentum, which constitutes a new result. Section~\ref{sec:static} is then devoted to the inclusion of static-field effects in the angular momentum loss, which are captured by the general formula obtained in Ref.~\cite{DiVecchia:2022owy}. In Section~\ref{sec:total} we combine the results of the previous sections to provide the complete result for the angular momentum carried by the gravitational field sourced by the two-body scattering and discuss its analytic continuation to the case of highly eccentric bound orbits. We discuss in detail recoil terms that arise when rewriting the covariant angular momentum tensor, on which we focus in this paper, in terms of the angular momentum vector. We also present several cross checks that validate our results with the existing literature and with an independent calculation that we performed in the PN limit. 
We collect our conventions on the kinematics in Appendix~\ref{app:Kin} and those concerning integrations and index contractions, including some useful Fourier transforms, in Appendix~\ref{app:FT}.

\section{General framework}
\label{sec:general}

\subsection{Eikonal operator}
\label{ssec:eikonal}
In the classical limit, the $S$-matrix simplifies and takes the form described by the eikonal operator, which can be constructed out of the following ingredients: 1) the conservative eikonal phase $2\tilde\delta$, whose derivative with respect to the impact parameter yields the impulse (see \cite{DiVecchia:2023frv} for a review) and which depends on spin already at tree level \cite{Vines:2017hyw,Guevara:2018wpp,Guevara:2019fsj}, 2) the waveform $\tilde{\mathcal{A}}^{\mu\nu}$ \cite{Kovacs:1977uw,Thorne:1980ru,Jakobsen:2020ksu,Mougiakakos:2020laz,Jakobsen:2021smu,Mougiakakos:2021ckm}, which at leading PM order is given by the Fourier transform (see Eq.~\eqref{A}) of the (connected) tree-level five-point amplitude $\mathcal{A}^{\mu\nu}$ and describes the emission of gravitational radiation, 3) the Weinberg soft factor augmented with the Feynman $-i0$ prescription, which encodes the effects due to the static Coulombic field \cite{Manohar:2022dea,DiVecchia:2022owy},
\begin{equation}\label{Fmunu}
	F^{\mu\nu}(k) = \Theta(\omega^\ast-k^0) \sum_{n} \frac{\sqrt{8\pi G} \,p_n^\mu p_n^\nu}{p_n\cdot k-i0}\,,
\end{equation}
with $\Theta$ the Heaviside step function and $\omega^\ast$ a cutoff to be sent to zero at a later stage.
Introducing graviton creation/annihilation operators $\hat a_k^\dagger$,  $\hat a_k$ (see Appendix~\ref{app:FT} for more details), the eikonal operator we are going to employ in this work takes the schematic form \footnote{See Refs.~\cite{DiVecchia:2022owy,DiVecchia:2022piu,DiVecchia:2023frv} for more details on the eikonal operator and on the inclusion of static effects.}
\begin{equation}\label{eikope4d}
	\hat{\mathcal{S}}
	\!=\!
	e^{i2\tilde\delta}
	e^{\int_k\left( F(k) a^\dagger_k-F^\ast(k) a_k \right)}
	e^{i\!\int_k\left( \tilde{\mathcal{A}}(k) \hat{a}^\dagger_k+\tilde{\mathcal{A}}^\ast\!(k) \hat{a}_k \right)}\,.
\end{equation}
The expectation values 
\begin{align}\label{trueexp}
	P^\alpha
	\!=\!
	\langle \Psi_\text{in} | \hat{\mathcal{S}}^\dagger \hat P^\alpha \hat{\mathcal{S}} | \Psi_\text{in}\rangle,\quad
	J^{\alpha\beta}
	\!=\!
	\langle \Psi_\text{in} | \hat{\mathcal{S}}^\dagger \hat J^{\alpha\beta} \hat{\mathcal{S}} | \Psi_\text{in}\rangle\,,
\end{align}
where $|\Psi_\text{in}\rangle$ denotes the initial state with no gravitons,
yield the energy-momentum and angular momentum carried by gravitational field  in the final state. Following \cite{DiVecchia:2022piu,DiVecchia:2023frv}, we shall distinguish between radiative quantities, denoted by bold symbols, and static ones, denoted by calligraphic symbols. Static contributions are absent for the emitted energy-momentum, so that $P^\alpha= \boldsymbol{P}^\alpha$. Conversely, for the angular momentum \cite{Damour:2020tta,Veneziano:2022zwh,DiVecchia:2022owy,DiVecchia:2022piu,Javadinezhad:2022ldc,Riva:2023xxm,Compere:2023qoa} we have $J^{\alpha\beta}= \boldsymbol{J}^{\alpha\beta}+\mathcal J^{\alpha\beta}$, with $\boldsymbol{J}^{\alpha\beta}$ the radiative piece and $\mathcal J^{\alpha\beta}$ the static piece \cite{Mougiakakos:2021ckm,Riva:2021vnj,Manohar:2022dea,DiVecchia:2022owy,DiVecchia:2022piu}.

\subsection{Spin effects in the five-point amplitude.}
The $2\to3$ amplitude in the classical limit $\mathcal{A}^{\mu\nu}$ for graviton emissions up to quadratic order in spin can be obtained, at tree level, from the stress-energy tensors $t^{\mu\nu}$ calculated in \cite{Mougiakakos:2021ckm,Riva:2021vnj,Riva:2022fru} via $\mathcal{A}^{\mu\nu}=4 (8\pi G)^{3/2} m_1^2 m_2^2 t^{\mu\nu} / (q_1^2 q_2^2)$. Here we shall focus for simplicity on the spin-orbit effects and consider the five-point amplitude $\mathcal A^{\mu\nu} =\mathcal A^{\mu\nu}_{0} + \mathcal A^{\mu\nu}_{s_1}$ , that includes terms up to linear order in the \emph{mass-rescaled} spin tensor of (say) particle number one, $s_1^{\mu\nu}$ (the general case of course can be obtained by symmetrizing over particle labels).  Let us recall that, in the classical limit, the non-spinning amplitude $\mathcal A^{\mu\nu}_0$ scales like $1/q^2$, while the spinning one $\mathcal A_{s_1}^{\mu\nu}$ scales like $1/q$, in agreement with the fact that $s_1^{\mu\nu}$ is itself of order $1/q$ in the classical limit. 
Crucially, both $\mathcal A_0^{\mu\nu}$ and $\mathcal A_{s_1}^{\mu\nu}$ vanish on shell when contracted with the graviton momentum $k_\mu$, a manifestation of gauge invariance. Finally, following again Ref.~\cite{Riva:2022fru}, we employ the covariant spin supplementary condition \cite{Vines:2017hyw,Guevara:2018wpp,Bern:2020buy}
\begin{equation}\label{ssc}
	s_{1\mu\nu} u_1^\nu= 0\,,
\end{equation}
where $u_1^\mu$ is the four-velocity of particle 1 in the incoming state.

\section{Radiative modes.}
\label{sec:radiative}
\subsection{Radiated energy-momentum}
\label{ssec:energymomentum}
From the expression of the eikonal operator \eqref{eikope4d}, one obtains that emitted linear momentum \eqref{trueexp} is expressed in terms of the leading-order waveform by \cite{Herrmann:2021lqe,Herrmann:2021tct,DiVecchia:2021bdo,Manohar:2022dea}
\begin{equation}\label{Prad}
	\boldsymbol{P}^\alpha = \int_{ k} \tilde{\mathcal{A}}\, k^\alpha \tilde{\mathcal{A}}^\ast\,,
\end{equation}
which can be recast as the Fourier transform of an integral over the 3-particle phase-space,
\begin{equation}\label{}
	\boldsymbol{P}^\mu = \operatorname{FT} \mathbb P^\mu\,,
\end{equation}
with the ``kernel'' (we adopt the conventions that quantities appearing on the right of the cut, depicted in red e.g. in Eq.~\eqref{3pcqspacesimpl} below, are subject to complex conjugation)
\begin{equation}\label{3pcqspacesimpl}
 \mathbb P^\alpha
	=\!
	\!\int\!
	d(\text{LIPS})\
	k^\alpha\!\!\!\!\!\!
	\begin{gathered}
		\begin{tikzpicture}[scale=.4]
			\draw[<-] (-4.8,5.17)--(-4.2,5.17);
			\draw[<-] (-1,5.15)--(-1.6,5.15);
			\draw[<-] (-1,3.15)--(-1.6,3.15);
			\draw[<-] (-1,.85)--(-1.6,.85);
			\draw[<-] (-4.8,.83)--(-4.2,.83);
			\draw[<-] (-2.85,1.7)--(-2.85,2.4);
			\draw[<-] (-2.85,4.3)--(-2.85,3.6);
			\path [draw, thick, blue] (-5,5)--(-3,5)--(-1,5);
			\path [draw, thick, color=green!60!black] (-5,1)--(-3,1)--(-1,1);
			\path [draw] (-3,3)--(-1,3);
			\path [draw] (-3,1)--(-3,5);
			\draw[dashed] (-3,3) ellipse (1.3 and 2.3);
			\node at (-1,3)[below]{$k$};
			\node at (-5,5)[left]{$p_1$};
			\node at (-5,1)[left]{$p_2$};
			\node at (-2.8,4)[left]{$q_1$};
			\draw[<-] (3.35,5.17)--(2.75,5.17);
			\draw[<-] (-.45,5.15)--(.15,5.15);
			\draw[<-] (-.45,3.15)--(.15,3.15);
			\draw[<-] (-.45,.85)--(.15,.85);
			\draw[<-] (3.35,.83)--(2.75,.83);
			\draw[<-] (1.4,1.7)--(1.4,2.4);
			\draw[<-] (1.4,4.3)--(1.4,3.6);
			\path [draw, thick, red] (-.7,.6)--(-.7,5.4);
			\path [draw, thick, blue] (3.55,5)--(1.55,5)--(-.45,5);
			\path [draw, thick, color=green!60!black] (3.55,1)--(1.55,1)--(-.45,1);
			\path [draw] (1.55,3)--(-.45,3);
			\path [draw] (1.55,1)--(1.55,5);
			\draw[dashed] (1.55,3) ellipse (1.3 and 2.3);
			\node at (1.35,4)[right]{$q-q_1$};
		\end{tikzpicture}
	\end{gathered}
\end{equation}
The Lorentz-invariant three-particle phase space measure in the soft limit takes the form
\begin{align}\label{measure3p}
	d(\text{LIPS})
	&=\tfrac{d^Dk}{(2\pi)^D}2\pi\theta(k^0)\delta(k^2)\\
	&
	\times
	\tfrac{d^Dq_1}{(2\pi)^D}2\pi\delta(2p_1\cdot q_1)
	2\pi\delta(2p_2\cdot (q_1 + k)).
	\nonumber
\end{align}
(see~Appendix~\ref{app:FT} for more details on the notation). Integrals of this type can be conveniently evaluated by means of reverse unitarity \cite{Anastasiou:2002qz,Anastasiou:2002yz,Anastasiou:2003yy,Herrmann:2021lqe,Herrmann:2021tct}. One first rewrites the phase-space delta functions as ``cut'' propagators, via the identity
\begin{equation}\label{}
	2i\pi\delta(x) = \frac{1}{x-i0}-\frac{1}{x+i0}\,,
\end{equation}
and after this step the resulting integrals can be handled with the \texttt{Mathematica} package \texttt{LiteRed} \cite{Lee:2012cn, Lee:2013mka} like conventional two-loop integrals in the soft region, see e.g.~\cite{DiVecchia:2020ymx,DiVecchia:2021bdo,Herrmann:2021tct}. We refer in particular to \cite[Sect.~3, 6.1]{DiVecchia:2021bdo} for more details about the integration and for the explicit expressions of the master integrals.

Let us now introduce the two projectors \cite{Jakobsen:2022fcj,Riva:2022fru}
\begin{align}
	\label{projectorSigma}
	\Pi^{\mu\nu} 
	&= \Sigma^{\mu\nu} - q^\mu q^\nu /q^2\,,
	\\ 
	\label{projectorPi}
	\Sigma^{\mu\nu} 
	&= \eta^{\mu\nu} + u_1^\mu \check u_1^\nu + u_2^\mu \check u_2^\nu\,,
\end{align}
which obey
\begin{equation}\label{orthogonals}
	\Sigma\indices{^\mu_\nu}u_{1,2}^\nu= 0\,,\qquad
	\Pi\indices{^\mu_\nu}u_{1,2}^\nu= 0\,,\qquad
	\Pi\indices{^\mu_\nu}q^\nu=0\,.
\end{equation}
In order to reduce the calculation of $\mathbb P^{\mu}$ to scalar integrals, we decompose it in terms of $\check u_1^\mu$, $\check u_2^\mu$, $q^\mu$, and of its component orthogonal to these vectors,
\begin{equation}\label{bbPdecomp}
\mathbb P^\mu = -(u_1\cdot \mathbb P)\,\check u_1^\mu
-(u_2\cdot \mathbb P)\,\check u_2^\mu
+(q\cdot \mathbb P) q^\mu/q^2 + \Pi\indices{^\mu_\nu}\mathbb P^\nu\,.
\end{equation} 
This reduces the problem to the calculation of the coefficients in Eq.~\eqref{bbPdecomp}. Due to the presence of the additional tensor $s_1^{\mu\nu}$ besides the particles' velocities and four momenta, the tensor reduction of the resulting integrals is not automatized by \texttt{LiteRed}. 
Therefore, we need to perform a suitable Passarino--Veltman decomposition of the integrand. To this end, for $u_{1,2}\cdot \mathbb P$ and $q\cdot \mathbb P$ we can employ
\begin{align}
	\begin{split}\label{PassVelt2}
s_{1\mu\nu} \ell_1^\mu \ell_2^\nu 
\to 
&-(\check u_2\cdot \ell_1)(s_{1\mu\nu}u_2^\mu q^\nu)\frac{\ell_2\cdot q}{q^2}\\
&
-\frac{\ell_1\cdot q}{q^2}(s_{1\mu\nu}q^\mu u_2^\nu)(\check u_2\cdot \ell_2)
	\end{split}
\\
\begin{split}\label{PassVelt1}
s_{1\mu\nu} \ell_i^\mu v^\nu 
\to 
&-(\check u_2\cdot \ell_i)(s_{1\mu\nu}u_2^\mu v^\nu)\\
&+\frac{\ell_i\cdot q}{q^2}(s_{1\mu\nu}q^\mu v^\nu),
\end{split}
\end{align}
where $\ell_{i=1,2}$ are the two loop momenta $\ell_1= -q_1$, $\ell_2= -k$ and $v^\mu \in \{u_1^\mu, u_2^\mu, q^\mu\}$.
For $\Pi\indices{^\mu_\nu}\mathbb P^\nu$, we have instead
\begin{widetext}
\begin{equation}\label{PassVeltPi2}
		\Pi\indices{^\alpha_\beta} \ell_i^\beta
		(s_{1\mu\nu} \ell_1^\mu \ell_2^\nu)
		\to 
		(\Pi_{\rho\sigma}\ell_i^\rho\ell_1^\sigma)\Pi^{\alpha\beta}s_{1\beta\nu}
		\left(-u_2^\nu (\ell_2\cdot \check u_2)+ q^\nu\frac{\ell_2\cdot q}{q^2}\right)
		+
		\left(-u_2^\nu (\ell_1\cdot \check u_2)+ q^\nu\frac{\ell_1\cdot q}{q^2}\right)
		\Pi^{\alpha\beta}s_{1\nu\beta}
		(\Pi_{\rho\sigma}\ell_i^\rho\ell_2^\sigma)\\
\end{equation}
\end{widetext}
for terms cubic in $\ell_{1,2}$ and
\begin{align}\label{PassVeltPi1}
	\Pi\indices{^\alpha_\beta} \ell_i^\beta
	(s_{1\mu\nu} \ell_j^\mu v^\nu) \to (\Pi_{\rho\sigma}\ell_i^\rho \ell_j^\sigma) \Pi^{\alpha\beta} s_{1\beta \nu} v^\nu
\end{align}
for those quadratic in $\ell_{1,2}$. In agreement with \cite{Riva:2022fru}, focusing on terms linear in spin, we find that $q\cdot \mathbb P$ vanishes, while, as expected by power counting, 
\begin{align}\label{}
	\label{u12bbP}
	-u_{1,2}\cdot \mathbb P &= \mathbb P_{u_{1,2}} s_{1\alpha\beta}u_2^\alpha q^\beta \sqrt{q^2}\,,
	\\
	\label{PibbP}
	\Pi\indices{^\mu_\nu} \mathbb P^\nu &= \mathbb P_\perp \Sigma^{\mu\alpha} s_{1\alpha\beta} q^\beta \sqrt{q^2}
\end{align}
where $\mathbb P_{u_{1,2}}$ and $\mathbb P_\perp$ are $q$-independent functions. Indeed, since $\mathcal A_{0}^{\mu\nu}$ and $\mathcal A_{s_1}^{\mu\nu}$ scale like $1/q^2$ and $1/q$ respectively, $k^\alpha$ scales like $q$ and the measure \eqref{measure3p} like $q^{4}$ in $D=4$, then \eqref{u12bbP} and \eqref{PibbP} must scale like $q^2$. Note that we were able to replace $\Pi^{\mu\alpha}$ with $\Sigma^{\mu\alpha}$ in \eqref{PibbP} owing to the antisymmetry of $s_{1\alpha\beta}$.  
The Fourier transforms of \eqref{u12bbP}, \eqref{PibbP} are then easily evaluated via \eqref{theFT1} and we find
\begin{align}\label{u12boldP}
	-u_{1,2}\cdot \boldsymbol P_{s_1} &= -\frac{3i}{2\pi b^5} \mathbb P_{u_{1,2}} s_{1\alpha\beta}u_2^\alpha b^\beta\,,\\
	\label{PiboldP}
	\Pi\indices{^\mu_\nu} \boldsymbol P_{s_1}^\nu &= -\frac{3i}{2\pi b^5} \mathbb P_\perp \zeta^\mu s_{1\alpha\beta}\zeta^\alpha b^\beta\,,
\end{align}
where we have introduced unit the vector
\begin{equation}\label{zeta}
	\zeta_\mu = -\frac{1}{b p_\infty} \, \epsilon_{\mu\alpha\beta\gamma} u_1^\alpha u_2^\beta b^\gamma
\end{equation}
orthogonal to $u_{1,2}$ and to $b$. We use the convention $\epsilon_{0123}=+1$ so that $\zeta^\mu$ is aligned with the relative orbital angular momentum vector of the incoming state, as one can easily check by evaluating \eqref{zeta} explicitly e.g.~in the rest frame of particle 2. In \eqref{PiboldP} we used that
\begin{equation}\label{SigmaDecomp}
	\Sigma^{\mu\alpha} = \frac{b^\mu b^\alpha}{b^2}+ \zeta^\mu \zeta^\alpha
\end{equation}
and dropped the $b^\mu b^\alpha/b^2$ term by exploiting the antisymmetry of $s_{1\alpha\beta}$. Note that in this case $\Pi\indices{^\mu_\nu}\mathbb P^{\nu}$ simply becomes proportional to $\zeta^\mu$ after performing the Fourier transform. The situation will be more complicated for the analogous angular momentum components.

Proceeding as described above, we obtain the following result for the spin-orbit contributions to the radiated energy-momentum, which reproduces the one obtained in Ref.~\cite{Riva:2022fru}, 
\begin{equation}\label{Ps1}
	\boldsymbol{P}^\alpha_{s_1}
	=
	-
	R_f
\left[
	\left(c_{u_1} \check{u}_1^\alpha
	+
	c_{u_2} \check{u}_2^\alpha\right)
	s_{1\mu\nu} u_2^\mu b^\nu
+
	c_\zeta
	\zeta^\alpha
	s_{1\mu\nu} \zeta^\mu b^\nu
\right],
\end{equation}
where  
\begin{equation}\label{Radfactor}
	R_f = \frac{\pi G^3 m_1^2 m_2^2}{b^5}
\end{equation}
and
\begin{equation}\label{}
	c_X = 3\,\frac{f_1^X}{p_1} + \frac{f_2^X}{p_2} \log\frac{\sigma+1}{2}+f_3^X\frac{ \operatorname{arccosh}\sigma}{p_3}\,,
\end{equation}
(here $X\in\{u_1,u_2,\zeta\}$)
with
\begin{equation}\label{p1p2p3}
\begin{split}
p_1 &= 48 (\sigma^2-1)^{3/2}(\sigma+1)\,,\\
p_2 &= -8 \sqrt{\sigma^2-1}(\sigma+1)\,,\\
p_3 &= \frac{16}{\sigma}(\sigma^2-1)^2\,.	
\end{split}
\end{equation}
The functions $f_1^X$, $f_2^X$, $f_3^X$
are given in Table~\ref{tab:Pradiative}
and depend on the relative Lorentz factor $\sigma = -u_1\cdot u_2$.
The expression \eqref{Ps1} is also provided in the ancillary file.
The comparison with Ref.~\cite{Riva:2022fru} is facilitated by Eq.~\eqref{convertspin} below (we also recall that a metric with opposite signature is used in that reference).

\begin{table}[b]%The best place to locate the table environment is directly after its first reference in text
	\fbox{
		\begin{minipage}{.9\linewidth}
				\vspace{-6pt}
			\begin{align*}
				f^{u_1}_1 &\!=\! 210 \sigma^6\!-\!356 \sigma^5\!-\!111 \sigma^4\!-\!1627 \sigma^3
				\!+\!5393 \sigma^2\!-\!4741 \sigma \!+\!1352
				\\
				f^{u_1}_2 &\!=\! 3 \left(35 \sigma ^4+115 \sigma ^3-135 \sigma ^2+49 \sigma -16\right) \\
				f^{u_1}_3 &\!=\! 15 \sigma  \left(2 \sigma ^2-3\right) \left(7 \sigma ^2-3\right) \\
				f^{u_2}_1 &\!=\! 210 \sigma^6 \!-\! 279 \sigma^5 \!-\! 219 \sigma^4 \!-\! 1350 \sigma^3 \!+\! 4732 \sigma^2\!-\!4243 \sigma \!+\!1245
				\\
				f^{u_2}_2 &\!=\! 12 \left(7 \sigma ^4+22 \sigma ^3-28 \sigma ^2+10 \sigma -3\right) \\
				f^{u_2}_3 &\!=\! 12 \sigma  \left(2 \sigma ^2-3\right) \left(7 \sigma ^2-3\right) \\
				f^{\zeta}_1 &\!=\! -(425 \sigma ^5-1215 \sigma ^4+2491 \sigma ^3-3957 \sigma ^2+2992 \sigma -760)
				\\
				f^{\zeta}_2 &\!=\! 3 \left(28 \sigma ^4+153 \sigma ^3-247 \sigma ^2+107 \sigma -25\right) \\
				f^{\zeta}_3 &\!=\! 3 \left(2 \sigma ^2-3\right) \left(28 \sigma ^3-15 \sigma ^2-12 \sigma +3\right)
			\end{align*}
		\end{minipage}
	}
	\caption{\label{tab:Pradiative}%
		Polynomials appearing in the emitted energy-momentum due to spin-orbit effects \cite{Riva:2022fru}.}
\end{table}
Eq.~\eqref{Ps1} holds for the case in which particle 1 carries spin and particle 2 is non spinning. The general case can be of course obtained by exploiting particle-interchange symmetry (we recall that $b^\mu$ changes sign under this transformation). 

\subsection{Radiated angular momentum}
\label{ssec:angularmomentum}
Turning to the radiated angular momentum $\boldsymbol J^{\alpha\beta}$, one can evaluate the corresponding expectation value \eqref{trueexp} by means of the eikonal operator \eqref{eikope4d} and show that its radiative piece takes the form  \cite{Manohar:2022dea,DiVecchia:2022owy,DiVecchia:2022piu}
\begin{equation}\label{}
	\boldsymbol J_{\alpha\beta} = \boldsymbol J^{(o)}_{\alpha\beta} + \boldsymbol J^{(s)}_{\alpha\beta}
\end{equation}
with
\begin{equation}\label{Jgravexpl}
i\boldsymbol J^{(o)}_{\alpha\beta}
=
\int_{{k}}
k_{[\alpha}
\frac{\partial\tilde{\mathcal{A}}}{\partial k^{\beta]}}\tilde{\mathcal{A}}^\ast
\,,\quad
\boldsymbol J^{(s)}_{\alpha\beta}
=
i \int_{{k}}
2\tilde{\mathcal{A}}^{\mu}_{[\alpha} \tilde{\mathcal{A}}_{\beta]\mu}^\ast\,,
\end{equation}
in terms of the waveform $\tilde{\mathcal{A}}^{\mu\nu}$.
The two terms in \eqref{Jgravexpl} are reminiscent of the graviton's orbital and spin angular momentum contributions, although actually only their sum is gauge invariant and physically meaningful.

For the ``spin'' term $\boldsymbol{J}^{(s)}_{\alpha\beta}$ we can apply the same strategy as before and reduce its calculation to the Fourier transform of a three-particle cut, with suitable index contractions and antisymmetrizations. For the ``orbital'' term $\boldsymbol{J}^{(o)}_{\alpha\beta}$, instead, we can apply the  formula derived in \cite{DiVecchia:2022piu}, which takes into account derivatives acting on the phase-space delta functions and holds in a frame where $b_1^\mu=b^\mu$ and $b_2^\mu=0$.
Using this strategy, like before, we can write the full $\boldsymbol{J}_{\alpha\beta}$ as the Fourier transform of a momentum-space kernel $\mathbb{J}_{\alpha\beta}$,
\begin{equation}\label{}
	\boldsymbol J_{\alpha\beta}= \operatorname{FT} \mathbb J_{\alpha\beta}\,,
	\qquad
	\mathbb J_{\alpha\beta} = \mathbb J^{(o)}_{\alpha\beta} + \mathbb J^{(s)}_{\alpha\beta}
\end{equation}
where
\begin{align}
	%\begin{split}
	&i\mathbb {J}^{(o)}_{\alpha\beta}
	\!=\!
	\!\int
	\!\! k_{[\alpha}
	\frac{\partial }{\partial k^{\beta]}}
	\!\!	\left[ d(\text{LIPS})\!\!\!\!\!\!
	\!\!\!\!\!\!
	\begin{gathered}
		\begin{tikzpicture}[scale=.4]
			\draw[<-] (-4.8,5.17)--(-4.2,5.17);
			\draw[<-] (-1,5.15)--(-1.6,5.15);
			\draw[<-] (-1,3.15)--(-1.6,3.15);
			\draw[<-] (-1,.85)--(-1.6,.85);
			\draw[<-] (-4.8,.83)--(-4.2,.83);
			\draw[<-] (-2.85,1.7)--(-2.85,2.4);
			\draw[<-] (-2.85,4.3)--(-2.85,3.6);
			\path [draw, thick, blue] (-5,5)--(-3,5)--(-1,5);
			\path [draw, thick, color=green!60!black] (-5,1)--(-3,1)--(-1,1);
			\path [draw] (-3,3)--(-1,3);
			\path [draw] (-3,1)--(-3,5);
			\draw[dashed] (-3,3) ellipse (1.3 and 2.3);
			\node at (-1,3)[below]{$k$};
			\node at (-5,5)[left]{$p_1$};
			\node at (-5,1)[left]{$p_2$};
			\node at (-2.8,4)[left]{$q_1$};
		\end{tikzpicture}
	\end{gathered}
	\!\!
	\right]
	\!\!\!
	\begin{gathered}
		\begin{tikzpicture}[scale=.4]
			\path [draw, thick, red] (.8,.6)--(.8,5.4);
			\draw[<-] (4.8,5.17)--(4.2,5.17);
			\draw[<-] (1,5.15)--(1.6,5.15);
			\draw[<-] (1,3.15)--(1.6,3.15);
			\draw[<-] (1,.85)--(1.6,.85);
			\draw[<-] (4.8,.83)--(4.2,.83);
			\draw[<-] (2.85,1.7)--(2.85,2.4);
			\draw[<-] (2.85,4.3)--(2.85,3.6);
			\path [draw, thick, blue] (5,5)--(3,5)--(1,5);
			\path [draw, thick, color=green!60!black] (5,1)--(3,1)--(1,1);
			\path [draw] (3,3)--(1,3);
			\path [draw] (3,1)--(3,5);
			\draw[dashed] (3,3) ellipse (1.3 and 2.3);
			\node at (2.8,4)[right]{$q-q_1$};
		\end{tikzpicture}
	\end{gathered}
	\nonumber
	\\
	\label{LL1Lp}
	&
	-\!
	u_{2[\alpha}
	\!
	\frac{\partial}{\partial q_{\parallel2}}
	\!\int\!\!  d(\text{LIPS})
	k^{\phantom{(2)}}_{\beta]}\!\!\!\!\!\!\!\!\!\!\!\!\!\!\!\!\!
	\begin{gathered}
		\begin{tikzpicture}[scale=.4]
			\draw[<-] (-4.8,5.17)--(-4.2,5.17);
			\draw[<-] (-1,5.15)--(-1.6,5.15);
			\draw[<-] (-1,3.15)--(-1.6,3.15);
			\draw[<-] (-1,.85)--(-1.6,.85);
			\draw[<-] (-4.8,.83)--(-4.2,.83);
			\draw[<-] (-2.85,1.7)--(-2.85,2.4);
			\draw[<-] (-2.85,4.3)--(-2.85,3.6);
			\path [draw, thick, blue] (-5,5)--(-3,5)--(-1,5);
			\path [draw, thick, color=green!60!black] (-5,1)--(-3,1)--(-1,1);
			\path [draw] (-3,3)--(-1,3);
			\path [draw] (-3,1)--(-3,5);
			\draw[dashed] (-3,3) ellipse (1.3 and 2.3);
			\node at (-1,3)[below]{$k$};
			\node at (-5,5)[left]{$p_1$};
			\node at (-5,1)[left]{$p_2$};
			\node at (-2.8,4)[left]{$q_1$};
			\draw[<-] (3.35,5.17)--(2.75,5.17);
			\draw[<-] (-.45,5.15)--(.15,5.15);
			\draw[<-] (-.45,3.15)--(.15,3.15);
			\draw[<-] (-.45,.85)--(.15,.85);
			\draw[<-] (3.35,.83)--(2.75,.83);
			\draw[<-] (1.4,1.7)--(1.4,2.4);
			\draw[<-] (1.4,4.3)--(1.4,3.6);
			\path [draw, thick, red] (-.7,.6)--(-.7,5.4);
			\path [draw, thick, blue] (3.55,5)--(1.55,5)--(-.45,5);
			\path [draw, thick, color=green!60!black] (3.55,1)--(1.55,1)--(-.45,1);
			\path [draw] (1.55,3)--(-.45,3);
			\path [draw] (1.55,1)--(1.55,5);
			\draw[dashed] (1.55,3) ellipse (1.3 and 2.3);
			\node at (1.35,4)[right]{$q-q_1$};
		\end{tikzpicture}
	\end{gathered}
	%\end{split}
\end{align}
with $q_{\parallel2}=-u_2\cdot q^\mu$,
and
\begin{equation}\label{LLspin}
	\mathbb {J}^{(s)}_{\alpha\beta}
	\!=\!
	2i
	\!\int\!\!
	d(\text{LIPS})
	\!\!
	\left[ 
	\begin{gathered}
		\begin{tikzpicture}[scale=.4]
			\draw[<-] (-4.8,5.17)--(-4.2,5.17);
			\draw[<-] (-1,5.15)--(-1.6,5.15);
			\draw[<-] (-1,3.15)--(-1.6,3.15);
			\draw[<-] (-1,.85)--(-1.6,.85);
			\draw[<-] (-4.8,.83)--(-4.2,.83);
			\draw[<-] (-2.85,1.7)--(-2.85,2.4);
			\draw[<-] (-2.85,4.3)--(-2.85,3.6);
			\path [draw, thick, blue] (-5,5)--(-3,5)--(-1,5);
			\path [draw, thick, color=green!60!black] (-5,1)--(-3,1)--(-1,1);
			\path [draw] (-3,3)--(-1,3);
			\path [draw] (-3,1)--(-3,5);
			\draw[dashed] (-3,3) ellipse (1.3 and 2.3);
			\node at (-1,3)[below]{$k$};
			\node at (-2.8,4)[left]{$q_1$};
		\end{tikzpicture}
	\end{gathered}
	\!\!
	\right]\indices{^\mu_{[\alpha}}\!
	\begin{gathered}
		\begin{tikzpicture}[scale=.4]
			\path [draw, thick, red] (0,.6)--(0,5.4);
		\end{tikzpicture}
	\end{gathered}
	\left[
	\begin{gathered}
		\begin{tikzpicture}[scale=.4]
			\draw[<-] (4.8,5.17)--(4.2,5.17);
			\draw[<-] (1,5.15)--(1.6,5.15);
			\draw[<-] (1,3.15)--(1.6,3.15);
			\draw[<-] (1,.85)--(1.6,.85);
			\draw[<-] (4.8,.83)--(4.2,.83);
			\draw[<-] (2.85,1.7)--(2.85,2.4);
			\draw[<-] (2.85,4.3)--(2.85,3.6);
			\path [draw, thick, blue] (5,5)--(3,5)--(1,5);
			\path [draw, thick, color=green!60!black] (5,1)--(3,1)--(1,1);
			\path [draw] (3,3)--(1,3);
			\path [draw] (3,1)--(3,5);
			\draw[dashed] (3,3) ellipse (1.3 and 2.3);
			\node at (2.8,4)[right]{$q-q_1$};
		\end{tikzpicture}
	\end{gathered}
\right]_{\beta]\mu}
\end{equation}
The calculation of the cut in the second line of \eqref{LL1Lp} must be performed by keeping $q_{\parallel_2}$ generic (while one can safely set $u_1\cdot q=0$), and only plugging in $q_{\parallel_2}=0$ \emph{after} taking the derivative with respect to it. Like in similar calculations performed in Refs.~\cite{DiVecchia:2022piu,Heissenberg:2022tsn}, we find that, when taken seprately, the two lines of \eqref{LL1Lp} have very singular small-velocity limits and would lead to an imaginary part in the final result; however, such singularities and imaginary parts cancel in the sum, leaving behind the well-behaved (real) result for the radiated angular momentum discussed below. These cancellations provide internal cross-checks for this type of calculations.

We can decompose $\mathbb J_{\alpha\beta}$ by considering its contractions with $u_{1,2}^\alpha$, $q^\alpha$ and $\Pi\indices{^\mu_\alpha}$, like we did for $\mathbb P^\alpha$ in \eqref{bbPdecomp}.
In order to reduce the calculation of such contractions to scalar integrals, we  again perform the Passarino--Veltman decomposition \eqref{PassVelt2}, \eqref{PassVelt1},
\eqref{PassVeltPi2}, \eqref{PassVeltPi1}.
We find that the nonzero components take the following form,
\begin{equation}\label{JmunucoefficientsBB}
	\begin{split}
	\mathbb J^{\mu\nu} 
	&= \mathbb J_{u_1q} (s_{1\alpha\beta} u_2^\alpha q^\beta)
	\check u_1^{[\mu}q^{\nu]}/\sqrt{q^2}\\
	&
	+
	\mathbb J_{u_2q} (s_{1\alpha\beta} u_2^\alpha q^\beta)
	\check u_2^{[\mu}q^{\nu]}/\sqrt{q^2}\\
	&
	+\mathbb J_{\perp u_1} s_{1\alpha\beta}u_2^{\alpha} \Pi^{\beta[\nu}\check u_1^{\mu]} \sqrt{q^2}\\
	&
	+\mathbb J_{\perp u_2} s_{2\alpha\beta}u_2^{\alpha} \Pi^{\beta[\nu}\check u_2^{\mu]} \sqrt{q^2}
	\\
	&
	+\mathbb{J}_{\perp q}\, q^\alpha s_{1\alpha\beta}\, \Pi^{\beta[\nu}\,q^{\mu]}/\sqrt{q^2}\,,
	\end{split}
\end{equation}
where $\mathbb{J}_{u_{1,2}q}$, $\mathbb J_{\perp u_{1,2}}$ and $\mathbb{J}_{\perp q}$ are $q$-independent functions, so that all terms scale like $q$, in agreement with power counting. Contrary to the situation that we encountered for instance in \eqref{PibbP}, however, now we cannot neglect the distinction between the two projectors $\Pi^{\beta\nu}$ and $\Sigma^{\beta\nu}$ in \eqref{projectorPi} and \eqref{projectorSigma}, and it is convenient to rewrite the former in terms of the latter, which is $q$-independent.
The Fourier transform to impact-parameter space of Eq.~\eqref{JmunucoefficientsBB} can  be evaluated with the help of \eqref{theFT0}, \eqref{theFT2} and induces a nontrivial reshuffling of the first four lines, leading to
\begin{equation}\label{}
	\begin{split}
		\boldsymbol J_{s_1}^{\mu\nu} 
		&=\frac{1}{\pi b^3} 
		\Big[-(\mathbb J_{u_1q}-\tfrac12 \mathbb J_{\perp u_1}) (s_{1\alpha\beta} u_2^\alpha b^\beta)
		\check u_1^{[\mu}b^{\nu]}/{b^2}\\
		&
		-
		(\mathbb J_{u_2 q} -\tfrac12 \mathbb J_{\perp u_2}) (s_{1\alpha\beta} u_2^\alpha b^\beta)
		\check u_2^{[\mu} b^{\nu]}/{b^2}\\
		&
		+(\tfrac12\mathbb J_{u_1q}- \mathbb J_{\perp u_1}) (s_{1\alpha\beta}u_2^{\alpha}\zeta^\beta) \check u_1^{[\mu}\zeta^{\nu]} 
		\\
		&
		+(\tfrac12\mathbb J_{u_2q}- \mathbb J_{\perp u_2}) (s_{1\alpha\beta}u_2^{\alpha}\zeta^\beta) \check u_2^{[\mu}\zeta^{\nu]} 
		\\
		&
		-\tfrac12 \mathbb{J}_{\perp q} (s_{1\alpha\beta} \zeta^\alpha b^\beta)\zeta^{[\mu}b^{\nu]}/{b^2}\Big],
	\end{split}
\end{equation}
where we also used the identity \eqref{SigmaDecomp} to rewrite $\Sigma^{\mu\nu}$ in terms of $b^\mu$ and $\zeta^\mu$.

Proceeding in this way, we obtain the following new result for the radiated angular momentum due to spin-orbit effects in a translation frame where $b_1^\alpha=b$, $b_2^\alpha=0$,
\begin{equation}\label{Js1}
	\begin{split}
\boldsymbol{J}_{s_1}^{\alpha\beta}
&=
R_f
\Big[
\left(
c_{\check u_1 b} \frac{u_1^{[\alpha} b^{\beta]}}{\sigma^2-1}
+
c_{\check u_2 b} \frac{u_2^{[\alpha} b^{\beta]}}{\sigma^2-1}
\right) s_{1\rho\sigma} u_2^\rho b^\sigma
\\
&+
\left(
c_{\check u_1 \zeta} \frac{u_1^{[\alpha} \zeta^{\beta]}}{\sigma^2-1}
+
c_{\check u_2 \zeta} \frac{u_2^{[\alpha} \zeta^{\beta]}}{\sigma^2-1}
\right) b^2 s_{1\rho\sigma} \zeta^\rho u_2^\sigma
\\
&
-
c_{b\zeta}\, b^{[\alpha}\zeta^{\beta]} s_{1\mu\nu} \zeta^\mu b^\nu
\Big]
	\end{split}
\end{equation}
where $R_f$ is given in \eqref{Radfactor} and
\begin{equation}\label{}
	c_{XY} = \frac{f_1^{XY}}{p_1} + \frac{f_2^{XY}}{p_2} \log\frac{\sigma+1}{2}+f_3^{XY}\frac{ \operatorname{arccosh}\sigma}{p_3}\,,
\end{equation}
(here $X,Y\in\{\check u_1,\check u_2,\zeta\}$)
while $p_1$, $p_2$, $p_3$ are given in \eqref{p1p2p3}. The functions $f^{XY}_1$, $f^{XY}_{2}$, $f^{XY}_3$ are listed in Table~\ref{tab:Jradiative}.
The expression for the result \eqref{Js1} is also provided in electronic format in the ancillary file.
\begin{table}[b]%The best place to locate the table environment is directly after its first reference in text
	\fbox{
		\begin{minipage}{.9\linewidth}
			\vspace{-6pt}
			\begin{align*}
				f^{\check u_1 b}_1 &\!=\! 2 (\sigma +1) (168 \sigma ^6-693 \sigma ^5+507 \sigma ^4+1014 \sigma ^3-2758 \sigma ^2\\
				&+2799 \sigma -797)
				\\
				f^{\check u_1 b}_2 &\!=\! -8 \left(14 \sigma ^5-31 \sigma ^4-66 \sigma ^3-68 \sigma ^2+104 \sigma -17\right) \\
				f^{\check u_1 b}_3 &\!=\! -8 \left(2 \sigma ^2-3\right) \left(14 \sigma ^4-30 \sigma ^3-21 \sigma ^2+18 \sigma -1\right) \\
				f^{\check u_2 b}_1 &\!=\! 150 \sigma ^6+1503 \sigma ^5-7635 \sigma ^4+21539 \sigma ^3-28603 \sigma ^2\\
				&+14482 \sigma -2612
				\\
				f^{\check u_2 b}_2 &\!=\! 210 \sigma ^5-194 \sigma ^4-1289 \sigma ^3-105 \sigma ^2+835 \sigma -113 \\
				f^{\check u_2 b}_3 &\!=\! \left(2 \sigma ^2-3\right) \left(210 \sigma ^4-404 \sigma ^3-225 \sigma ^2+228 \sigma -5\right) \\
				f^{\check u_1 \zeta}_1 &\!=\! -(84 \sigma ^7+597 \sigma ^6-417 \sigma ^5-1746 \sigma ^4+3274 \sigma ^3\\
				&-4199 \sigma ^2+4259 \sigma -2620)
				\\
				f^{\check u_1 \zeta}_2 &\!=\! -4 \left(14 \sigma ^5-31 \sigma ^4-66 \sigma ^3-68 \sigma ^2+92 \sigma -29\right) \\
				f^{\check u_1 \zeta}_3 &\!=\! -4 \left(2 \sigma ^2-3\right) \left(14 \sigma ^4-30 \sigma ^3-21 \sigma ^2+18 \sigma -13\right) \\
				f^{\check u_2 \zeta}_1 &\!=\! 390 \sigma ^6+429 \sigma ^5-2643 \sigma ^4+4795 \sigma ^3\\
				&-5927 \sigma ^2+4532 \sigma -2416
				\\
				f^{\check u_2 \zeta}_2 &\!=\!105 \sigma ^5-55 \sigma ^4-415 \sigma ^3-423 \sigma ^2+530 \sigma -142 \\
				f^{\check u_2 \zeta}_3 &\!=\! \left(2 \sigma ^2-3\right) \left(105 \sigma ^4-160 \sigma ^3-135 \sigma ^2+96 \sigma -46\right) \\
				f^{b \zeta}_1 &\!=\! -3 (249 \sigma ^5-501 \sigma ^4+1026 \sigma ^3-1842 \sigma ^2+1205 \sigma -233 )
				\\
				f^{b\zeta}_2 &\!=\! 12 \left(7 \sigma ^4+22 \sigma ^3-28 \sigma ^2+10 \sigma -3\right) \\
				f^{b\zeta}_3 &\!=\! 12 \sigma  \left(2 \sigma ^2-3\right) \left(7 \sigma ^2-3\right)
			\end{align*}
		\end{minipage}
	}
	\caption{\label{tab:Jradiative}%
		Polynomials appearing in the emitted angular momentum due to spin-orbit effects.}
\end{table}

Eq.~\eqref{Js1} provides the terms in the radiated angular momentum linear in $s_{1}^{\mu\nu}$ in a frame where $b_1^\alpha=b$, $b_2^\alpha=0$. Therefore, interchanging all particle labels in it one obtains the terms linear in $s_2^{\mu\nu}$ in a frame where $b_1^\alpha=0$, $b_2^\alpha=-b^\alpha$ instead. Of course, one can obtain the radiated angular momentum in any desired translation frame by using the following  transformation law 
under a translation \eqref{translation}, 
\cite{Manohar:2022dea}
\begin{equation}
	\label{translation-J}
	\boldsymbol{J}^{\alpha\beta}\to \boldsymbol J^{\alpha\beta}+a^{[\alpha} \boldsymbol{P}^{\beta]}\,.
\end{equation} 
and the explicit form \eqref{Ps1} for $\boldsymbol{P}^\alpha_{s_1}$.

It can be convenient to introduce the spin vector $s_1^\mu$ as well, which is linked to the spin tensor $s_1^{\mu\nu}$ by
\begin{subequations}
\begin{align}\label{}
	s_1^\rho &= - \frac12 \epsilon^{\rho\alpha\beta\gamma} u_{1\alpha}s_{1\beta\gamma}\,,\\
	s_1^{\mu\nu} &= \epsilon^{\mu\nu\rho\sigma} u_{1\rho} s_{1\sigma}\,,
\end{align}
\end{subequations}
so that in particular, letting
(note that $p_\infty$ is a dimensionless variable, unlike $p$ introduced in Section~\ref{sec:total})
\begin{equation}\label{}
	\sigma= \sqrt{1+p_\infty^2}\,,
\end{equation}
one has
\begin{subequations}\label{convertspin}
\begin{align}\label{}
	s_{1\alpha\beta} u_2^\alpha b^\beta &= b p_\infty (s_1\cdot \zeta)\,,\\
	s_{1\alpha\beta} \zeta^\alpha u_2^\beta &= p_\infty (s_1\cdot b)/b\,,\\
	s_{1\alpha\beta} b^\alpha \zeta^\beta&= b (s_1\cdot e)\,, 
\end{align}
\end{subequations}
where we defined the unit vector
\begin{equation}\label{vectore}
	e^\mu = (u_2^\mu - \sigma u_1^\mu)/p_\infty = - p_\infty \check u_2^\mu\,. 
\end{equation}

The result in Eq.~\eqref{Js1} holds in a reference frame whose origin is such that $b_1^\alpha=b$, $b_2^\alpha=0$.
Going to a frame where $b_1^\alpha = 0$ and $b_2^\alpha = -b^\alpha$, by performing a translation \eqref{translation-J} with $a^\alpha= -b^\alpha$, let us denote
\begin{equation}\label{}
	\widehat{\boldsymbol{J}}_{s_1}^{\alpha\beta} 
	=
	\frac{1}{b^2 R_f}
	\left(
	\boldsymbol{J}^{\alpha\beta}_{s_1} 
	-
	b^{[\alpha}\boldsymbol{P}_{s_1}^{\beta]}
	\right)
\end{equation}
and list the approximate expressions for its components in the Post-Newtonian (PN) limit $p_\infty\to0$. For the spatial components, i.e.~the radiated angular momentum proper, we find
\begin{subequations}\label{spatJPN}
\begin{align}
	\frac{b_\alpha}{b} \widehat{\boldsymbol{J}}^{\alpha\beta}_{s_1} e_\beta
	&= \frac{67}{5}\, p_\infty s_1\cdot\zeta  +\mathcal O(p_\infty^3),\\
	\zeta_\alpha \widehat{\boldsymbol{J}}^{\alpha\beta}_{s_1} e_\beta
	&= \frac{23}{2}\, p_\infty (s_1\cdot b)/b  +\mathcal O(p_\infty^3),\\
	\frac{b_\alpha}{b} \widehat{\boldsymbol{J}}^{\alpha\beta}_{s_1} \zeta_{\beta}
	&= \frac{43}{10}\, p_\infty s_1\cdot e  +\mathcal O(p_\infty^3)\,,
\end{align}
\end{subequations}
and for the mixed space-time components, i.e. the radiated mass dipole, we find
\begin{subequations}\label{timeJPN}
\begin{align}
	-\frac{b_\alpha}{b} \widehat{\boldsymbol{J}}^{\alpha\beta}_{s_1} u_{1\beta}
	&= \frac{2}{15}\, s_1\cdot\zeta  +\mathcal O(p_\infty^2),\\
	-\zeta_\alpha \widehat{\boldsymbol{J}}^{\alpha\beta}_{s_1} u_{1\beta}
	&= -\frac{7}{30}\, (s_1\cdot b)/b  +\mathcal O(p_\infty^2)\\
	e_\alpha \widehat{\boldsymbol{J}}^{\alpha\beta} u_{1\beta}&=0
	\,.
\end{align}
\end{subequations}
Of  course the complete Poincar\'e covariant result \eqref{Js1} provides the resummation of the PN expansions \eqref{spatJPN}, \eqref{timeJPN} to all orders in the velocity to leading order in the coupling, $\mathcal O(G^3)$.
Conversely, as we shall discuss in Section~\ref{sec:total}, independent PN calculations of the leading-order components \eqref{spatJPN}, \eqref{timeJPN} will provide useful cross checks of the results obtained so far.

\section{Static Modes}
\label{sec:static}
The results discussed in the previous section are obtained from the appropriate ``square'' of a \emph{connected} $2\to3$ amplitude $\mathcal A^{\mu\nu}$, which involves gravitons with strictly non-zero frequency. However, unlike for the loss of energy-momentum, zero-frequency contribution arise from the eikonal operator \eqref{eikope4d} when inserted in the expectation value for the angular momentum  loss \eqref{trueexp}, and depend on the ``Weinberg factor'' $F^{\mu\nu}$ in \eqref{Fmunu}. These capture the effects due to static gravitational fields, and have been discussed in detail in Refs.~\cite{DiVecchia:2022owy,DiVecchia:2022piu,DiVecchia:2023frv}. To include their contributions to the angular momentum loss, we can employ Eq.~(3.30) of \cite{DiVecchia:2022owy}, which computes them for a generic gravitational process and which evaluates to \cite{DiVecchia:2022owy,DiVecchia:2023frv,Bini:2022wrq}
\begin{equation}\label{Jabnm}
	\mathcal J^{\alpha\beta} = - \frac{G}{2} (p_1-p_2)^{[\alpha}Q^{\beta]}\mathcal I(\sigma)+\mathcal O(G^4)\,,
\end{equation}
where  
\begin{equation}\label{}
	Q^\alpha= p_1^\alpha+p_4^\alpha = -p_2^\alpha-p_3^\alpha
\end{equation}
is the conservative impulse which includes $\mathcal O(G)$ and $\mathcal O(G^2)$ contributions, and 
\begin{equation}\label{}
	\tfrac12\,\mathcal{I}(\sigma)=\frac{8-5\sigma^2}{3(\sigma^2-1)}
	+
	\frac{\sigma(2\sigma^2-3)\operatorname{arccosh}\sigma}{(\sigma^2-1)^{3/2}}
\end{equation}
is the function also appearing in $\mathcal O(G^3)$ radiation-reaction effects \cite{Damour:2020tta,DiVecchia:2021ndb}.
Like the leading soft theorem from which it was obtained, the formula \eqref{Jabnm} only depends on the momenta of the hard particles, and to obtain explicit expressions it is sufficient to substitute the PM expansion of the impulse $Q$. 
Physically, it captures the loss of angular momentum due to the change in the Coulombic fields sourced by the massive particles before and after the scattering, or, at a diagrammatic level, ``zero-frequency gravitons'' attached to the massive lines, as is clear from the fact that the cutoff $\omega^\ast$ has been sent to zero.
Including the impulse up to $\mathcal O(G^2)$ and up to linear order in spin, one has
\begin{equation}\label{}
	Q^\mu= Q_\text{1PM}^\mu + Q_{\text{1PM},s_1}^\mu +
	Q_\text{2PM}^\mu + Q_{\text{2PM},s_1}^\mu +\cdots
\end{equation}
and therefore the static spin-orbit effects in the angular momentum read
\begin{equation}\label{Js1stat}
	\begin{split}
		\mathcal J_{s_1}^{\alpha\beta} &= - \frac{G}{2} (p_1-p_2)^{[\alpha}(Q_{\text{1PM},s_1}+Q_{\text{2PM},s_1})^{\beta]}\mathcal I(\sigma)\\
		&+\mathcal O(G^4)
	\end{split}
\end{equation}
We have checked the $\mathcal O(G^2)$ term of Eq.~\eqref{Js1stat} against Eq.~(4.9) of Ref.~\cite{Alessio:2022kwv}, which provides the loss of spatial angular momentum in the center-of-mass frame to all orders in the spin. The $\mathcal O(G^3)$ of Eq.~\eqref{Js1stat} is instead new.

For completeness, let us quote the leading and subleading order impulse to linear order in spin from Ref.~\cite{Jakobsen:2022fcj}. We collect the relevant expression for generic spin orientation in the ancillary file, while only reproducing here the expression for the aligned-spin configuration, in which case $s_1^\mu$ is aligned with $\zeta^\mu$ and thus
\begin{equation}\label{aligned}
	s_1\cdot \zeta= a_1>0
\end{equation}
coincides with the magnitude of the spin vector.
Then, to leading PM order,
\begin{equation}\label{}
	Q_{\text{1PM},s_1}^\mu= 
	\frac{4\sigma G m_1 m_2}{b^3} b^\mu a_1
\end{equation}
and to subleading PM order
\begin{widetext}
\begin{equation}\label{}
	Q_{\text{2PM},s_1}^{\mu}
	=
	\frac{G^2 m_1 m_2}{2b^3(\sigma^2-1)}
	\Big[
	\pi  \left(4 m_1+3 m_2\right) \sigma  \left(5 \sigma ^2-3\right)  a_1 \frac{b^\mu}{b}
	+
	\frac{16 \sigma (2\sigma^2-1)}{\sqrt{\sigma^2-1}}
	\left(
	(\sigma m_1+m_2)u_1^\mu
	-
	(\sigma m_2+m_1)u_2^\mu
	\right)
	a_1
	\Big].
\end{equation}
\end{widetext}

\section{Total angular momentum loss}
\label{sec:total}

\subsection{Complete result and analytic continuation}
\label{ssec:complete}
In this section, we combine the results obtained in the previous ones in order to provide the complete expression for the angular momentum carried by the gravitational field, including both radiative and static contributions. We shall also discuss its analytic continuation from unbound to highly eccentric bound trajectories, focusing on aligned-spin configurations.

To this end, let us introduce the four-velocity of an observer at rest with respect to the \emph{incoming} center of mass,
\begin{equation}\label{}
	t^\mu =  -\frac{p_1^\mu+p_2^\mu}{\sqrt{-(p_1+p_2)^2}} = \frac{m_1 u_1^\mu + m_2 u_2^\mu}{E}\,,
\end{equation}
with 
\begin{equation}\label{}
	E =\sqrt{-(p_1+p_2)^2}= \sqrt{m_1^2+2 m_1 m_2 \sigma + m_2^2} 
\end{equation}
the total (incoming) center-of-mass energy,
and the spatial momentum of particle 1 as seen from that observer,
\begin{equation}\label{}
	\begin{split}
		p^\mu 
		&= - (p_1^\mu + t^\mu (p_1\cdot t)) \\
		&= 
		\frac{m_1 m_2}{E^2} \left(\left(m_1 \sigma +m_2\right) u_1^\mu-\left(m_2 \sigma +m_1\right) u_2^\mu \right),
	\end{split}
\end{equation}
which obeys
\begin{equation}\label{}
	p = \sqrt{p^2} = m_1 m_2\sqrt{\sigma^2-1}/E\,.
\end{equation}

Going to a frame where the center of mass (or ``center of energy'') sits in the origin of the transverse plane means to impose
\begin{equation}\label{CoE}
	E_1 b_1^\mu + E_1 b_2^\mu = 0
\end{equation}
with $E_{1,2}$ given by
\begin{equation}\label{}
	\qquad E_{1,2} = t\cdot p_{1,2}\,.
\end{equation}
Therefore, in order to map \eqref{Js1} to such a frame we need to perform a translation \eqref{translation-J} with
\begin{equation}\label{}
	a^\mu = - \frac{E_1}{E} b^\mu = - \frac{m_1(m_1+\sigma m_2) b^\mu}{m_1^2+2m_1m_2\sigma+m_2^2}\,, 
\end{equation}
thus introducing
\begin{equation}\label{Js1CM}
	\widetilde{\boldsymbol{J}}_{s_1}^{\alpha\beta} = \boldsymbol{J}_{s_1}^{\alpha\beta}
	-
	\frac{m_1(m_1+\sigma m_2) b^{[\alpha}\boldsymbol{P}^{\beta]}_{s_1}}{m_1^2+2m_1m_2\sigma+m_2^2}\,.
\end{equation}

Adding the radiative expression evaluated in the center-of-mass translation frame \eqref{Js1CM} and the static one \eqref{Js1stat}, which is invariant under translations, we obtain the complete angular momentum loss up to $\mathcal O(G^3)$ and up to linear order in spin, for generic spin alignments,
\begin{equation}\label{Js1totCM}
	J_{s_1}^{\alpha\beta}  =   \widetilde{\boldsymbol{J}}_{s_1}^{\alpha\beta}+\mathcal J_{s_1}^{\alpha\beta}\,.
\end{equation}
Eq.~\eqref{Js1totCM} thus provides the covariant angular momentum that is carried by the gravitational field resulting from the two-body encounter, or equivalently the total mechanical angular momentum loss experienced by the two-body system during the scattering, expressed in terms of the angular momentum tensor, and is valid in the center-of-mass translation frame.
Let us also define 
\begin{equation}\label{bJpDef}
	J_{s_1} =\frac{b_\alpha}{b}  J_{s_1}^{\alpha\beta}\frac{p_\beta}{p}
	=
	J^{\mathcal O(G^2)}_{s_1}
	+
	J^{\mathcal O(G^3)}_{s_1}
	+
	\mathcal O(G^4)\,.
\end{equation}

Let us now deduce from \eqref{bJpDef} an expression valid for  bound orbits in the large eccentricity limit, by following the Boundary-to-Bound  map \cite{Kalin:2019inp,Kalin:2019rwq,Saketh:2021sri,Cho:2021arx} for the aligned-spin case \eqref{aligned}. The first step is to express $J_{s_1}$ as a function of the variables $L=bp$, $a_1$, $\sigma$ and look at its analytic properties as a function of (complex) $\sigma$. The objective is to analytically continue the result, obtained for energies $E = m\sqrt{1+2\nu(\sigma-1)}>m$ that is, $\sigma>1$, to the bound case, in which $E<m$, that is, $\sigma<1$.
We note that the functions $\log(\frac{\sigma+1}{2})$ and $\frac{\operatorname{arccosh}\sigma}{\sqrt{\sigma^2-1}}=\frac{\operatorname{arccos}\sigma}{\sqrt{1-\sigma^2}}$ are analytic for $\operatorname{Re}\sigma>-1$, so they can be straightforwardly extended. 
The second step is to use the relation \cite{Saketh:2021sri,Cho:2021arx}
\begin{equation}\label{BtoB}
	J_{s_1,B} (L, a_1,\sigma) 
	=
	J_{s_1} (L, a_1,\sigma)
	+
	J_{s_1} (-L, -a_1,\sigma)\,.
\end{equation}
Like in the non-spinning case, the $\mathcal O(G^2)$ term $J_{s_1}^{\mathcal O(G^2)}$ is odd under the simultaneous sign change $L\to -L$, $a_1\to-a_1$ so that its contribution to \eqref{BtoB} vanishes. Conversely, the $\mathcal O(G^3)$ term $J_{s_1}^{\mathcal O(G^3)}$ is even and picks up a factor of two, so that
\begin{equation}\label{jBOUND}
	J_{s_1,B}(L, a_1,\sigma)  = 2 J_{s_1}^{\mathcal O(G^3)}(L, a_1,\sigma) 
\end{equation}
to this order.

Alternatively, one can sum over the branch choice one has to perform when analytically continuing $\sigma$ \cite{DiVecchia:2023frv}. It is easy to see that only $J_{s_1}^{\mathcal O(G^2)}(L, a_1,\sigma) $ is sensitive to this choice, due to an overall power of $\sqrt{\sigma^2-1}$ in its expression, which has a branch cut for $\sigma<1$, while $J_{s_1}^{\mathcal O(G^3)}(L, a_1,\sigma)$ is analytic for $\operatorname{Re}\sigma>-1$. As a result, one can define
\begin{equation}\label{BtoB2}
	J_{s_1,B} (L, a_1,\sigma) 
	=
	J_{s_1} (L, a_1,\sigma)_+
	+
	J_{s_1} (L, a_1,\sigma)_-
\end{equation}
where the subscript $\pm$ corresponds to choosing $\sqrt{\sigma^2-1}=\pm i \sqrt{1-\sigma^2}$,
reaching again the same conclusion \eqref{jBOUND}. Again, the same considerations also apply to the non-spinning expressions \cite{Manohar:2022dea,DiVecchia:2022piu,DiVecchia:2023frv}.

\subsection{Angular momentum vector and recoil}
The angular momentum carried by the gravitational field generated during the two-body encounter, $J_{s_1}^{\mu\nu}$, corresponds to a loss of mechaical angular momentum of the system comprising the two massive objects. Initially, the total mechanical angular momentum reads
(as usual, without loss of generality, we only include the spin of particle 1)
\begin{equation}\label{Jmech}
	J_\text{mech}^{\mu\nu} = L_\text{mech}^{\mu\nu} + m_1 s_1^{\mu\nu}
\end{equation}
with 
\begin{equation}\label{}
	L^{\mu\nu}_\text{mech}= - b_1^{[\mu} p_1^{\nu]} - b_2^{[\mu} p_1^{\nu]} = b^{[\mu} p^{\nu]}\,,
\end{equation}
where we used the condition \eqref{CoE}, and its variation must obey the balance law \cite{Manohar:2022dea,DiVecchia:2022piu}
\begin{equation}\label{JmDJtensor}
	J^{\mu\nu}=- \Delta J_\text{mech}^{\mu\nu}\,.
\end{equation}

It is often useful to introduce the mechanical angular momentum vector that is dual to $J_\text{mech}^{\mu\nu}$ with respect to the center-of-mass time direction $t^\mu$, via
\begin{equation}\label{JmuJbetagamma}
	J_\text{mech}^{\mu}  = - \frac{1}{2} \epsilon\indices{^\mu_{\alpha\beta\gamma}} t^{\alpha} J_\text{mech}^{\beta\gamma}
\end{equation}
and the angular momentum vector of the gravitational field via the corresponding variation,
\begin{equation}\label{JmDJvector}
	J^{\ast\mu}=- \Delta J_\text{mech}^{\mu}\,.
\end{equation}
Substituting \eqref{JmuJbetagamma} into \eqref{JmDJvector} and recalling \eqref{JmDJtensor}, we find 
\begin{equation}\label{JplusR}
	J^{\ast\mu}= J^\mu + R^\mu
\end{equation}
where
\begin{subequations}
\begin{align}
	\label{Jmueps}
	J^\mu &= - \frac{1}{2} \epsilon\indices{^\mu_{\alpha\beta\gamma}} t^{\alpha} J^{\beta\gamma} \,,\\
	\label{Rmueps}
	R^\mu &=  - \frac{1}{2} \epsilon\indices{^\mu_{\alpha\beta\gamma}} (- \Delta t^{\alpha}) J_\text{mech}^{\beta\gamma}\,.
\end{align} 
\end{subequations}
The vector $J^\mu$ is thus the dual of $J^{\mu\nu}$ as defined with respect to the time direction in the \emph{initial} center-of-mass frame, while $R^\mu$ encodes the recoil induced by the variation of the time direction itself, 
\begin{equation}\label{Deltat}
	\begin{split}
	\Delta t^\mu 
	&= - \frac{1}{E}\,[\boldsymbol{P}^\mu + t^\mu (t\cdot \boldsymbol{P})] \\
	&= - \frac{1}{E} \,\left(p^\mu\, \frac{p\cdot \boldsymbol{P}}{p^2} + \zeta^\mu \zeta\cdot \boldsymbol{P}\right).
	\end{split}
\end{equation}
Here we used that $\boldsymbol{P}\cdot b=0$ to up to linear order in spin.
From \eqref{Jmueps}, using the identity
\begin{equation}\label{}
	- \epsilon\indices{^\mu_{\alpha\beta\gamma}} t^\alpha b^{\beta} p^\gamma= (bp) \zeta^\mu
\end{equation}
and similar ones obtained by cyclic permutations of three of the basis vectors, one obtains
\begin{equation}\label{JmuExplicit}
	J^\mu = 
	\frac{1}{bp}
	\left[
	\left(b_\alpha J^{\alpha\beta} p_\beta\right) \zeta^\mu
	+
	\left(p_\alpha J^{\alpha\beta} \zeta_\beta\right) b^\mu
	+
	\left(\zeta_\alpha J^{\alpha\beta} b_\beta\right) {p^\mu}
	\right]
\end{equation}
and, combining \eqref{Rmueps} with \eqref{Jmech} and \eqref{Deltat}, the recoil term can be cast in the following form up to linear order in spin,
\begin{equation}\label{RecoilExplicit}
	\begin{split}
	R^\mu 
	&= 
	\left[
	\zeta^\mu (s_1\cdot \zeta)
	+
	\frac{b^\mu}{b^2}\,(s_1\cdot b)
	\right]
	\mathcal R_1\\
	&-
	t^\mu 
	\left[
	pb \, \mathcal R_2 - \frac{m_1}{p}\,(s_1\cdot e)\, \mathcal R_1
	\right]
	\end{split}
\end{equation}
with
\begin{equation}\label{}
	\mathcal R_1 = -(p\cdot \boldsymbol{P})/E\,,
	\qquad
	\mathcal R_2 = - (\zeta \cdot \boldsymbol{P})/E\,.
\end{equation}
For our purposes, since we work to leading order in spin, we can focus on the non-spinning contributions to $\mathcal R_1$ and on the linear-in-spin contribution to $\mathcal R_2$. Note also that the recoil $R^\mu$ vanishes altogether in the non-spinning case \cite{Manohar:2022dea}. In other words, recoil does not affect the change in mechanical (orbital) angular momentum vector in the absence of spin.

As we shall see in the next section, it is important to take into account both terms in \eqref{JplusR} when performing comparisons between the result \eqref{Js1totCM} and the expressions available in the literature and for checking the balance law in terms of angular momentum \emph{vector} (as opposed to \emph{tensor}). Combining \eqref{JmuExplicit} and \eqref{RecoilExplicit}, we obtain the final expressions
\begin{equation}\label{finalJast}
\begin{split}
	J^{\ast\mu}
	&=
	\zeta^\mu
	\left[
	\frac{1}{bp}\, (b_\alpha J^{\alpha\beta} p_\beta)+(\zeta\cdot s_1) \mathcal R_1
	\right]\\
	&+
	\frac{b^\mu}{b}
	\left[
	\frac{1}{p}\, (p_\alpha J^{\alpha\beta} \zeta_\beta)+ \frac{(b\cdot s_1)}{b} \mathcal R_1
	\right]\\
	&
	+
	\frac{p^\mu}{bp}\,(\zeta_\alpha J^{\alpha\beta} b_\beta)
	\\
	&
	-
	t^\mu 
	\left[
	pb \, \mathcal R_2 - \frac{m_1}{p}\,(s_1\cdot e)\, \mathcal R_1
	\right].
\end{split}
\end{equation}
We remark that $p^\mu$-component of $J^{\ast\mu}$ is unaffected by the recoil, while the $t^\mu$-component is simply given by a combination of recoil terms.
Note also that, by construction, $J^{\mu}$ and $J^{\ast\mu}$ do not include any information about the mass-dipole loss $t_\alpha J^{\alpha\mu}$, which  is instead available in \eqref{Js1totCM}.

\subsection{Consistency checks}
\label{ssec:consistency}
To obtain cross-checks of the present results, we performed several comparisons.
We first checked that Eq.~\eqref{Js1totCM} reproduces the result in Eq.~(5.22) of Ref.~\cite{Cho:2021arx} for the aligned-spin configuration in the PN expansion (these results were also recently reproduced in \cite{Bini:2023mdz} to leading PN order). Note that it is important to first translate from the canonical spin formulation employed in that reference to the covariant one employed here. To this end, it is sufficient to use Eq.~(2.3) of \cite{Cho:2021arx}; see Ref.~\cite{Riva:2022fru} for more details on this step. To $\mathcal O(G^2)$, only $\mathcal J_{s_1}^{\alpha\beta}$ contributes and one has
\begin{equation}\label{}
	J_{s_1}^{\mathcal O(G^2)}
	=
	-\frac{32 G^2m^3\nu^2}{b^2}\left(
	\frac{1}{5}p_\infty^3+\frac{7\nu-1}{70}\,p_\infty^5
	\right)a_1
	+\mathcal O(p_\infty^7)
\end{equation}
with 
\begin{equation}\label{}
	\nu= \frac{m_1m_2}{m^2}\,,\qquad m = m_1+m_2\,,
\end{equation}
in perfect agreement with Ref.~\cite{Cho:2021arx}.
To $\mathcal O(G^3)$, both $\mathcal J_{s_1}^{\alpha\beta}$ and $\boldsymbol J_{s_1}^{\alpha\beta}$ contribute and one finds
	\begin{align}
		\nonumber
		J_{s_1}^{\mathcal O(G^3)} 
		&= 
		\frac{\pi G^3m^4\nu^2}{b^3}
		\Bigg[
		\left(-\frac{287}{15}-\frac{2 \Delta }{3}\right) p_{\infty}
		\\
		\nonumber
		&+
		\left(-\frac{284}{21} -\frac{1723 \Delta }{840}+ \frac{2}{15}(122+\Delta) \nu \right) p_{\infty}^3
		\Bigg]a_1
		\\
		\label{SLPNPorto}
		&+
		\mathcal O(p_\infty^5)\,.
	\end{align}
with $\Delta= \sqrt{1-4\nu}$.
To compare the subleading PN accurate expression \eqref{SLPNPorto}  with Ref.~\cite{Cho:2021arx}, we need to take into account the recoil term dictated by Eq.~\eqref{finalJast}.  For the aligned-spin case only its first line contributes and, projecting along $\zeta^\mu$, yields
\begin{equation}\label{}
	J_{s_1}^\ast  = J_{s_1} + a_1 \mathcal R_1\,.
\end{equation}
Using that
\begin{equation}\label{}
	\mathcal R_1 = \frac{\pi G^3 m^4\nu^2 }{b^3}
	\left[
	\frac{37}{30} \nu\, p_{\infty}^3\, \Delta
	+\mathcal O(p_\infty^5)
	\right],
\end{equation}
we find agreement between $J_{s_1}^\ast=J_{s_1}+a_1 \mathcal{R}_1$, with $J_{s_1}$ as in Eq.~\eqref{SLPNPorto} above, and Eq.~(5.22) of Ref.~\cite{Cho:2021arx}.
This is to be expected because the approach adopted in that reference was to integrate the fluxes of Refs.~\cite{Porto:2010zg,Cho:2021mqw} expressed in terms of the source multipole moments evaluated in the instantaneous center-of-mass frame. Taking into account the recoil term is also important to obtain a match between Eq.~\eqref{jBOUND} above and Eq.~(5.21) of Ref.~\cite{Cho:2021arx} for the angular momentum lost during one period of elliptic-like motion in the case of a bound system \footnote{The relative factor of $1/2$ in \eqref{jBOUND} arises from the combination $\pi^{-1}\arccos(-e^{-1})$ in the large eccentricity limit $e\to\infty$ in Eq.~(5.22) of Ref.~\cite{Cho:2021arx}.}.

As a separate check, we performed an independent calculation of the angular momentum to leading order in the PN limit following the strategy described in Refs.~\cite{Manohar:2022dea,Heissenberg:2022tsn}, for generic spin alignments. We start from the tree-level $2\to3$ amplitude $\mathcal{A}^{\mu\nu}(k)$ and expand it for small $p_\infty$ in the scaling region $k^\alpha \sim\mathcal{O}(p_\infty)$ \cite{Kovacs:1977uw,Kovacs:1978eu,Mougiakakos:2021ckm}. We then calculate its Fourier transform \eqref{A} directly the PN expansion in a frame where $b_1^\alpha=0$. Finally, we substitute into the covariant expression for $\boldsymbol{J}^{\alpha\beta}$ \eqref{Jgravexpl} and perform the integration over $k$, without resorting to reverse unitarity. Thanks to the PN limit, this step only involves integrals of expressions quadratic in Bessel functions evaluated at the same argument, which are elementary and can be conveniently computed in \texttt{Mathematica}. Contracting \eqref{Jgravexpl} with the basis vectors $u_{1}^\alpha$, $e^\alpha$, $b^\alpha$, $\zeta^\alpha$, we thus reproduce the leading PN orders of all 6 independent components \eqref{spatJPN} and \eqref{timeJPN}, which depend on all 3 components of the spin vector, $s_1\cdot \zeta$, $s_1\cdot b$ and $s_1\cdot e$.
To arrive at \eqref{timeJPN}, when performing the small-velocity expansion, one can safely restrict to the leading non-spinning and linear-in-spin PN waveform $\mathcal{O}(1/p_\infty)$.  For \eqref{spatJPN}, instead, one must also include the next-to-leading $\mathcal{O}(p_\infty^0)$ corrections for both non-spinning and linear-in-spin contributions, and the final result arises from the interference terms. 

Finally, we compared with the loss of angular momentum vector $-\Delta J_\text{mech}^\mu$ in the center-of-mass frame for generic spin alignments obtained via a linear response approach in Ref.~\cite{Jakobsen:2023hig} (to be published simultaneously), finding perfect agreement for the full PM results at order $\mathcal O(G^3)$ for all four independent components of $J^{\ast\mu}$ listed in Eq.~\eqref{finalJast}. As already mentioned, since we are performing this comparison at the level of angular momentum vector, the inclusion of recoil terms as discussed above is crucial in order to correctly translate between the two results.
We refer to \cite{Jakobsen:2023hig} for more details on the linear response formula in the presence of spin which builds on \cite{Bini:2012ji,Bini:2021gat,Manohar:2022dea,Jakobsen:2022zsx,Bini:2022enm}.

\section{Conclusions}
In this paper, we calculated for the first time the contributions to the angular momentum loss due to spin-orbit effects up to $\mathcal O(G^3)$, using the eikonal operator. We presented the outcome of this calculation for generic spin alignments and in a fully Poincar\'e covariant way, in such a way that it can be easily translated to any desired reference frame.

It will be interesting to extend the results obtained here to quadratic order in spin, i.e.~spin-spin effects, as well, by using the waveforms or stress-energy tensors calculated in \cite{Jakobsen:2021lvp,Riva:2022fru}, and perhaps even beyond the quadratic-in-spin frontier using the EFT framework of Refs.~\cite{Aoude:2022trd,Aoude:2022thd,Bautista:2023szu,Aoude:2023vdk}, although fixing the Wilson coefficients of this EFT in general depending on the nature of the colliding objects represents an outstanding open problem \cite{Bautista:2021wfy,Bautista:2022wjf}. 

Similarly,  a natural problem to tackle consists in the calculation of $J^{\alpha\beta}$ to $\mathcal O(G^4)$, three loops on the amplitude side, even in the non-spinning case, and two concrete  steps towards this goal would be to derive a suitable generalization of the momentum-space kernels \eqref{LL1Lp}, \eqref{LLspin} and to add the non-linear memory contributions to Eq.~(3.30) of \cite{DiVecchia:2022owy}, which start to be relevant precisely at $\mathcal O(G^4)$.

It would also be interesting to apply a suitable analytic continuation and to perform the eccentricity resummation needed to provide results valid for the regime of quasicircular orbits \cite{Kalin:2019rwq,Kalin:2019inp,Saketh:2021sri,Cho:2021arx}, thus going beyond the analytic continuation for large eccentricity performed here for aligned spins \cite{Saketh:2021sri,Cho:2021arx,Heissenberg:2022tsn}, and to obtain the corresponding flux. In this connection, a relevant open problem consists in the extension of he Boundary-to-Bound map to generic spin alignments.

The linear response approach of Refs.~\cite{Bini:2012ji,Bini:2021gat,Bini:2022wrq} was adapted to the eikonal framework in \cite{DiVecchia:2022piu}, for the non-spinning case. It would be desirable to achieve the same in the presence of spin, building on the results of Ref.~\cite{Jakobsen:2023hig}, which may also facilitate the calculation of higher-order terms

Finally, a relevant step towards concrete phenomenological applications will be to study the impact of our results on waveform models \cite{Antonelli:2019ytb,Khalil:2022ylj}.

\begin{acknowledgments}
I would like to thank Paolo Di Vecchia, Gustav Jakobsen, Gustav Mogull, Jan Plefka, Rafael Porto, Massimiliano M. Riva and Rodolfo Russo for very useful discussions.
This work is supported by the Knut and Alice Wallenberg Foundation under grant KAW 2018.0116. Nordita is partially supported by Nordforsk.

\vspace{10pt}

\paragraph*{Note added.} While working on this project, we became aware of a closely related calculation performed via linear response by Gustav U. Jakobsen, Gustav Mogull, Jan Plefka and Benjamin Sauer \cite{Jakobsen:2023hig}. We thank them for sharing preliminary results that allowed us to check agreement with the present ones and for coordinating publication.

\end{acknowledgments}

\appendix

\section{Kinematics}
\label{app:Kin}
We employ an all-outgoing convention for the external momenta, so that $-p_1$ and $-p_2$ are the physical momenta of particle 1 and 2 in the far past, while $p_4$ and $p_3$ are their momenta in the far future. We use the mostly-plus convention according to which the Minkowski metric is $\eta_{\mu\nu}=\text{diag}(-,+,+,+)$, so that the mass-shell conditions are $p_1^2=-m_1^2$ and $p_2^2 = -m_2^2$ (and similarly for $p_3$, $p_4$).
The four-velocities of the incoming particles are defined by 
\begin{equation}\label{}
	u_{1}^\mu=-p_1^\mu/m_1\,,\qquad u_{2}^\mu=-p_2^\mu/m_2\,.
\end{equation}
We denote the relative Lorentz factor by
\begin{equation}\label{}
	\sigma = -u_1\cdot u_2=1/\sqrt{1-v^2}\,,
\end{equation} 
where $v$ is the spatial velocity of object 1 as measured in the rest frame of object 2 (or vice versa).
A related variable, which is particularly useful in the PN limit, is $p_\infty= \sqrt{\sigma^2-1}$.  
It is also convenient to define ``dual'' velocities by
\begin{equation}
\check u_1^\mu = \frac{\sigma u_2^\mu-u_1^\mu}{\sigma^2-1}\,,\qquad
\check u_2^\mu = \frac{\sigma u_1^\mu-u_2^\mu}{\sigma^2-1}\,,
\end{equation}
in such a way that $u_i\cdot \check u_j = -\delta_{ij}$. We also introduce a unit vector  $e^\mu$ in \eqref{vectore}, closely related to $\check u_2^\mu$. 

The unit vector $\zeta^\mu$ is defined in Eq.~\eqref{zeta} and is aligned with the relative orbital angular momentum of the incoming state.
We define $b^\mu= b_1^\mu-b_2^\mu$ to be the impact parameter, where for simplicity~$b_i\cdot u_j=0$ for $i,j=1,2$.
The symmetric mass ratio $\nu$ and the total mass $m$ of the system are defined by 
\begin{equation}\label{}
	\nu= m_1m_2/m^2\,,\qquad
	m=m_1+m_2
\end{equation}
while we let
\begin{equation}\label{}
	\Delta=(m_1-m_2)/m=\sqrt{1-4\nu}>0\,.
\end{equation}

\section{Integration and Index Contraction}
\label{app:FT}
We define the integral over the graviton's phase space as follows,
\begin{align}\label{kLIPS}
	\int_{k} &= \int \frac{d^Dk}{(2\pi)^D}\,2\pi\theta(k^0)\delta(k^2)\,.
\end{align}
Moreover, we define
\begin{equation}\label{Aa}
	\tilde{\mathcal{A}}(k)\,\hat a^\dagger_k=\sum_i \epsilon^{(i)}_{\mu\nu}(k)^\ast\tilde{\mathcal{A}}^{\mu\nu}(k)\, \hat{a}^\dagger_i(k)
\end{equation}
where $i=1,2$  runs over the two polarizations $\epsilon^{(i)}_{\mu\nu}(k)$ (we adopt an analogous notation for expressions involving $F^{\mu\nu}$). The graviton ladder operators $\hat a_i$, $\hat a_i^\dagger$ obey
\begin{equation}\label{}
	2\pi\theta(k^0)\delta(k^2)[\hat a_i(k),\hat a^\dagger_j(k')] \!=\! (2\pi)^{\!D}\delta^{(D)}(k-k')\delta_{ij}\,.
\end{equation}
We avoid displaying explicitly the index contractions involving $\mathcal A_{\mu\nu}$, letting for instance 
\begin{equation}\label{suppressedindices}
	\mathcal A\, \mathcal A' = \mathcal A_{\mu\nu}\, \mathcal A'^{\mu\nu}-\frac1{D-2}\,\mathcal A^{\mu}_\mu \mathcal A'^\nu_\nu,.
\end{equation}

We define 
\begin{equation}\label{FT2}
\operatorname{FT}\mathcal{M}\!= \!\!\int \!\!\!\tfrac{d^Dq}{(2\pi)^D}\,2\pi\delta(2p_1\cdot q)2\pi\delta(2p_2\cdot q)\,e^{ib\cdot q}\mathcal M(q)\,.
\end{equation}
Fourier transforms of this kind in which $\mathcal M(q)$ is proportional to a (complex) power of $q^2$ can be evaluated via (see e.g.~Appendix A.3 of \cite{DiVecchia:2023frv})
\begin{equation}\label{theFT}
	\operatorname{FT}[(q^2)^\nu] = \frac{2^{2\nu}}{\pi}\frac{\Gamma(1+\nu)}{\Gamma(-\nu)(b^2)^{1+\nu}}
\end{equation}
in four spacetime dimensions.
From \eqref{theFT} and its derivatives with respect to $b^\alpha$, we obtain in particular
\begin{align}
	\label{theFT0}
	\operatorname{FT}[\sqrt{q^2}] 
	&=- \frac{1}{2 \pi b^3}\,,\\
	\label{theFT1}
	\operatorname{FT}[q^\alpha \sqrt{q^2}] 
	&= -\frac{3i b^\alpha}{2\pi b^5}\,,\\
	\begin{split}
	\label{theFT2}
	\operatorname{FT}[q^\alpha q^\beta / \sqrt{q^2}] 
	&= \frac{1}{2\pi b^3}\left(
	\Sigma^{\alpha\beta}-\frac{3b^\alpha b^\beta}{b^2}
	\right)\\
	&= 
	-
	\frac{b^\alpha b^\beta}{\pi b^3}
	+
	\frac{\zeta^\alpha\zeta^\beta}{2\pi b^3}
	\,,
	\end{split}\end{align}
where $\Sigma^{\alpha\beta}$ is the projector orthogonal to $u_{1,2}$ defined in \eqref{projectorSigma}, $\zeta^\mu$ is the unit vector orthogonal to $u_{1,2}$, $b$ introduced in \eqref{zeta} and we used the identity \eqref{SigmaDecomp} in \eqref{theFT2}.

The $2\to3$ amplitude in the classical limit 
\begin{equation}\label{Aq1q2k}
	\mathcal{A}^{\mu\nu} (q_1,q_2,k)= 	\begin{gathered}
		\begin{tikzpicture}[scale=.4]
			\draw[<-] (-4.8,5.17)--(-4.2,5.17);
			\draw[<-] (-1,5.15)--(-1.6,5.15);
			\draw[<-] (-1,3.15)--(-1.6,3.15);
			\draw[<-] (-1,.85)--(-1.6,.85);
			\draw[<-] (-4.8,.83)--(-4.2,.83);
			\draw[<-] (-2.85,1.7)--(-2.85,2.4);
			\draw[<-] (-2.85,4.3)--(-2.85,3.6);
			\path [draw, thick, blue] (-5,5)--(-3,5)--(-1,5);
			\path [draw, thick, color=green!60!black] (-5,1)--(-3,1)--(-1,1);
			\path [draw] (-3,3)--(-1,3);
			\path [draw] (-3,1)--(-3,5);
			\draw[dashed] (-3,3) ellipse (1.3 and 2.3);
			\node at (-1,3)[below]{$k$};
			\node at (-5,5)[left]{$p_1$};
			\node at (-5,1)[left]{$p_2$};
			\node at (-2.8,4)[left]{$q_1$};
			\node at (-2.8,2)[left]{$q_2$};
		\end{tikzpicture}
	\end{gathered}
\end{equation}
(the drawing in Eq.~\eqref{Aq1q2k} is only meant as a visual help for recalling the definition of its arguments and does not represent an actual diagram topology)
is instead linked to the leading-order waveform $\tilde{\mathcal{A}}^{\mu\nu}(k)$ by the Fourier transform
\begin{equation}\label{A}
	\begin{split}
\tilde{\mathcal{A}}^{\mu\nu}(k) &= \!\!\int\!\!\! \frac{d^Dq_1}{(2\pi)^D}\,2\pi\delta(2p_1\cdot q_1)2\pi\delta(2p_2\cdot q_2)\\
&\times e^{ib_1\cdot q_1+ib_2\cdot q_2}\mathcal A^{\mu\nu}(q_1,q_2,k)\,,
	\end{split}
\end{equation}
with the momentum conservation condition
\begin{equation}\label{}
	q_1+q_2+k=0\,.
\end{equation}
Under translations,
\begin{equation}\label{translation}
	b^\mu_{1,2}\!\to\! b^\mu_{1,2}+a^\mu,\qquad
	\tilde{\mathcal{A}}^{\mu\nu}(k)\! \to \! e^{-ia\cdot k}\!\tilde{\mathcal{A}}^{\mu\nu}(k)\,,
\end{equation}
and, by \eqref{translation}, one immediately deduces the transformation property \eqref{translation-J}.
By \eqref{translation}, going to a frame where $b_2^\alpha= 0$, one has
\begin{align}\label{Awhenb2=0}
		&\tilde{\mathcal{A}}^{\mu\nu}(k) = \!\!\int\!\!\! \frac{d^Dq_1}{(2\pi)^D}\,2\pi\delta(2p_1\cdot q_1)\, e^{ib\cdot q_1}\\
		&\times 2\pi\delta(2p_2\cdot (q_1+k))\,\mathcal A^{\mu\nu}(q_1,-q_1-k,k)\,.\nonumber
\end{align}
A useful feature of the expression \eqref{Awhenb2=0} is that $k^\alpha$ only appears in its second line, so that the derivatives needed to calculate $\boldsymbol{J}^{(o)}_{\alpha\beta}$ in Eq.~\eqref{Jgravexpl} leave its first line untouched. 

%\nocite{*}

\providecommand{\href}[2]{#2}\begingroup\raggedright\endgroup

\end{document}